\documentclass[prd,aps,twocolumn,leterpaper,preprintnumbers,footinbib]{revtex4}
\usepackage[T2A]{fontenc}
\usepackage{amsmath,amssymb}
\usepackage{eucal}
\usepackage{color}
\usepackage[dviwindo]{graphicx}
\usepackage{epsfig,array,graphicx}
\usepackage{rotating,rotfloat}
\usepackage{multirow}
\usepackage{subfigure}

\newcommand{\bs}{\begin{subequations}}
\newcommand{\es}{\end{subequations}}
\numberwithin{equation}{section}
\newcommand{\ben}{\begin{eqnarray}}
\newcommand{\een}{\end{eqnarray}}
\newcommand{\la}{\label}

\begin{document}

\title{Withholding Potentials,  Absence of Ghosts  and Relationship\\ between Minimal Dilatonic Gravity and f(R) Theories}

\author{P. P.  Fiziev}
\affiliation{Department of Theoretical Physics, University of Sofia, Boulevard
5 James Bourchier, Sofia 1164, Bulgaria\\
and\\
BLTF, JINR, Dubna, 141980 Moscow Region, Rusia}

\email{fiziev@phys.uni-sofia.bg}
\email{fiziev@theor.jinr.ru}
\begin{abstract}

We study the relation between  Minimal Dilatonic Gravity (MDG) and f(R) theories of gravity
and establish strict conditions for their {\em global} equivalence.
Such equivalence takes place only for a certain class of cosmological potentials,
dubbed here {\em withholding potentials}, since they prevent change of the sign of dilaton $\Phi$.
The withholding property ensures the attractive character of gravity,
as well as absence of ghosts and a tachyon in the gravi-dilaton sector and yields certain asymptotic of the functions $f(R)$.
Large classes of withholding cosmological potentials and functions $f(R)$ are found and described in detail.
The particle content of the gravi-dilaton sector is found using perturbation theory around
de Sitter vacuum of MDG.
Two phenomena: scalaron waves and
induction of gravitational waves by the scalaron field are discussed
using the derived wave equations for MDG scalaron and graviton in the de Sitter background.
Seemingly, the MDG and f(R) theories, offer a unified description of dark energy and dark matter.
\vskip .3truecm
\noindent{PACS number(s): 04.50.Kd, 04.62.+v, 98.80.Jk}

\end{abstract}
\sloppy
\maketitle

\section{Introduction}\label{Intro}

Consider a scalar-tensor model of gravity  with action for the gravi-dilaton sector
\cite{OHanlon72,Fiziev00a,Fiziev00b,Fiziev02,Fiziev03,Landau,Fujii03}\footnote{The
sign conventions for the curvature tensor $R_{\alpha\beta, \gamma}{}^\delta$, Ricci
tensor $R_{\alpha\beta}$  and scalar curvature $R$ are
opposite to the ones used in articles \cite{Fiziev00a,Fiziev00b,Fiziev02,Fiziev03}
and reference \cite{Landau}. The signature of the metric is $\{+,-,-,-\}$.}:
\ben
{\cal A}_{g,\Phi}={\frac c {2\kappa}}\int d^4 x\sqrt{|g|}
 \bigl( \Phi R - 2 \Lambda U(\Phi) \bigr).
\la{A_MDG}
\een

We call this model {\em the minimal dilatonic gravity} (MDG)
It corresponds to the Branse-Dicke theory with identically vanishing parameter $\omega$.

Here $\kappa=8\pi G_{N}/c^2>0$ is the Einstein constant,
$\Lambda>0$ is the cosmological constant, and $\Phi\in (0,\infty)$ is the dilaton field. The values $\Phi$ must be positive
since a change of the sign of $\Phi$ entails a non-physical
change of the sign of the gravitational factor $G_N/\Phi$
and leads to antigravity.
Besides, the value $\Phi=0$ must be excluded since it leads to an
infinite gravitational factor and makes the Cauchy problem
not well posed \cite{EFarese}.
The value $\Phi=\infty$ turns off the gravity and is also physically unacceptable.

The scalar field $\Phi$ is introduced to consider {\em a variable} gravitational factor $G(\Phi)=G_N/\Phi$ instead
of the Newton constant $G_N$.
The cosmological potential $U(\Phi)$ is introduced to consider {\em a variable} cosmological factor
instead of the cosmological constant $\Lambda$. In general relativity (GR) with cosmological constant $\Lambda$
we have $\Phi\equiv 1$ and $U(1)\equiv 1$ .
Due to its specific physical meaning,
the scalar field $\Phi$ must have quite unusual properties.

The cosmological potential $U(\Phi)$ must be a positive {\em single valued} function
of the dilaton field $\Phi$ by astrophysical reasons.
Here we shall justify additional physical requirements
for the cosmological potential $U(\Phi)$
as a necessary ingredient of a sound MDG model.

Some physical and astrophysical consequences of MDG are described in
\cite{Fiziev00a,Fiziev00b,Fiziev02,Fiziev03,EFarese}.
In \cite{EFarese} a theory of cosmological perturbations for general scalar-tensor theories,
including MDG, was developed as a generalization of the approach of \cite{Boisseau00}.
For a recent attempt to develop a theory of cosmological perturbation
using action \eqref{A_MDG} see \cite{Bertacca11}.

It was shown \cite{Fiziev00a,Fiziev00b,Fiziev02,Fiziev03} that MDG can describe simultaneously:
1) The inflation and the graceful exit to the present day accelerating de Sitter expansion of the Universe.
2) The reconstruction of the cosmological potential $U(\Phi)$,
using the scalar factor $a(t)$ in the Friedman-Robertson-Walker model.
3) The way to avoid any conflicts with the existing solar system and laboratory gravitational experiments
using a large enough mass of the dilaton field $m_\Phi\gtrsim 10^{-3}\, \text{eV}/ c^2 $.
4) The time of inflation as a reciprocal quantity to the mass of dilaton $m_\Phi$.

In the last decade the $f(R)$ theories of gravity with the action
\ben
{\cal A}_{f(R)}={\frac c {2\kappa}}\int d^4 x\sqrt{|g|}f(R)
\la{Action_fR}
\een
attracted much attention, since they seem to offer a possible explanation
of the observed accelerating expansion of the Universe without introducing "dark energy"
and may be also an explanation of the observational problem of missing mass without introducing "dark matter",
see
\cite{Woodard06,Felice06,Faraoni06,Nojiri07,Nojiri08a,Cognola08,Nojiri08b,Felice10,Sotiriou10,Tsujikawa10,Nojiri11,Rubakov11,Clifton12,Bamba12,Gannouji12,Starobinsky80,Muiller88,Nojiri03,Carroll04,Nojiri04,Abdalla05,Capozzielo06a,Capozziello06b,Appleby07,Hu07,Starobinsky07,Amendola07,Li07,Bamba08,Tsujikawa08,Nojiri08c,Appleby09,Motohashi10} %
and a large amount of references therein. For $f(R)=R-2\Lambda$ we obtain once more GR with the cosmological constant.

Some of the popular choices of the function $f(R)$ are:
1. The Starobinsky 1980  model \cite{Starobinsky80};
2. The Carroll et al. 2004 model \cite{Carroll04};
3. The Capozziello et al. 2006 model \cite{Capozziello06b};
4. The Appleby et al. 2007 model \cite{Appleby07};
5. The Hu et al. 2007 model \cite{Hu07};
6. The Starobinsky 2007 model \cite{Starobinsky07,Motohashi10};
7. The Amendola-Li et al. 2007 model \cite{Amendola07}, \cite{Li07};
8. The Tsujikawa  2008 model \cite{Tsujikawa08};
9. The Appleby et al. 2009 model \cite{Appleby09};
10. The Gannouji et al. 2012 model \cite{Gannouji12}.

A few of the functions $\Delta f(R)=f(R)-R$, being carefully chosen, are cosmologically viable.
Their further modification is still under investigation as
a promising approach to the modern astrophysical problems.

Many authors think that using the Legendre transformation \cite{Arnold89,Zia09}
of the $f(R)$-action \eqref{Action_fR} to the MDG-action \eqref{A_MDG} and vice versa
they are able to prove the equivalence of these two generalizations of GR.
We show that a careful analysis leads to the opposite conclusion.
In general, the two models are equivalent only {\em locally} which
is not enough to ensure identical {\em physical} content.

Therefore in the present paper we study the {\em global} equivalence between MDG and  the $f(R)$ theories.
We derive a complete set of additional requirements and define the class of the cosmological potentials $U(\Phi)$
which indeed lead to {a global} equivalence and thus avoid some of the well-known  problems, like physically
unacceptable singularities, ghosts, etc., in the $f(R)$ theories.

Some of the above authors consider the last problems seriously already
if not explicitly but implicitly\footnote{The author is grateful
to the unknown referee for this remark.}.
Hence, these problems need further investigation.

A part of the necessary additional requirements are known and dispersed in a large amount of
the existing literature on the $f(R)$ theories,
where the problem of the global equivalence with MDG has been never discussed.
One of the goals of the present paper is to collect all results at one place,
to represent their strict derivation and correct usage,
thus making transparent the mathematical structure and the physical content of the theory.

An obvious inequality of the MDG and $f(R)$ theories lies in the fact that we do not have a physical intuition on how
to choose the function $\Delta f(R)$. In contrast, our large experience in different kinds of physical theories,
starting with classical mechanics, is very helpful in the construction of the cosmological potential $U(\Phi)$
\cite{Fiziev00a,Fiziev00b,Fiziev02,Fiziev03}.

In Section \ref{FieldEqs} we present the different forms of the field equations
of MDG and a discussion of their specific features and peculiarities.

In Section \ref{Legendre} we consider the specific Legendre transform between
MDG and $f(R)$ theories  and the conditions for the global equivalence of these two models.

In Section \ref{WithholdingProperty} we
introduce a new type of withholding potentials $V(\Phi)$, $U(\Phi)$ and functions $f(R)$.
The main objective of this paper is to find a large enough sets of such objects
to be prepared to describe the real problems.

In Section \ref{Ghosts} we demonstrate the solution of the ghost problem
in both the MDG and $f(R)$ theories using the witholding property.

The form of a withholding self-interaction potential $\tilde V(\varphi)$
in the Einstein frame is presented in Section \ref{Gosts_fR}.

Another presented result is a correct formulation of the perturbation theory around de Sitter vacuum
and clarification of the particle content of the gravi-dilaton sector in MDG
(Section \ref{ParticleContent}).

In the Concluding Remarks (Section \ref{conclusion}) we outline the basic results and some
open physical problems.

\section{The field equations of MDG}\label{FieldEqs}

The variation of the action \eqref{A_MDG}  with respect to the dilaton field $\Phi$ gives
{\em without} any restrictions of the variation $\delta\Phi$
on the boundary $\partial V^4$ the {\em algebraic} relation
\ben
R = 2 \Lambda U_{,\Phi}(\Phi).
\la{R_Phi}
\een

A decisive feature of MDG is the existence of such a relation,
instead of a differential equation for the field $\Phi$.
According to \eqref{R_Phi}, the dilaton field $\Phi$ has no space-time properties and evolution
independent of the space-time scalar curvature $R$. In particular:

i) In space-times with $R=\text{const}$ due to relation \eqref{R_Phi} we have $\Phi=\text{const}$
and gravitational factor $G(\Phi)=\text{const}$.  This corresponds to
our  physical expectations for gravity, for example, in a flat space-time with $R=0$.

ii) In the case of validity of the cosmological principle (CP) at very large scales ($\sim 10^2 \div 10^3$ Mpc)
the 3D space is homogeneous and isotropic, maybe even flat. Then all quantities must depend only on the cosmic time $t$.
As a result of CP, we have $R=R(t)$. Then \eqref{R_Phi} yields $\Phi=\Phi(t)$, i.e.
the gravitational factor also obeys the CP.

In any other scalar-tensor model, or in any other field frame,
instead of the algebraic equation \eqref{R_Phi}
we will have a partial differential equation for the dilaton field
$\Phi(x)$.
Thus, we would lose the above attractive properties of MDG which
seem to correspond to the physical reality with good precision,
if one interprets the dilaton field $\Phi(x)$
as a local strength scale of gravitational interaction.
For example, it would be quite strange to have some dynamics of the local intensity
of gravitational interaction in an empty flat space-time.

The variation of the action \eqref{A_MDG}  with respect to the metric field $g_{\alpha\beta}$ is more complicated.
Taking into account the identity \eqref{x} and the four restrictions \eqref{restric_dg}
on the surface $\partial V^{(3)}$ (see Appendix A) we obtain
the vacuum equations for gravy-dilaton sector in the form
\ben
\Phi G_{\alpha\beta}+ \Lambda U(\Phi)g_{\alpha\beta}+\nabla_\alpha\nabla_\beta\Phi-g_{\alpha\beta}\Box \Phi=0.
\la{MDG_Eq}
\een

The trace of the eqs. \eqref{MDG_Eq} yields the
dynamical equation in vacuum:
\ben
\Box \Phi+\Lambda V_{,\Phi}(\Phi)=0.
\la{Phi_eq}
\een
By construction it is {\em written} in terms of the dilaton field $\Phi$.
Here $V_{,\Phi}(\Phi)={2\over 3}\left(\Phi U_{,\Phi}- 2U\right)={2\over 3}\Phi^3{d\over {d\Phi}}\left(\Phi^{-2}U\right)$
is the derivative of the dilatonic potential. For convenience, we use the normalization $V(1)=0$. Then
\ben
V(\Phi)={2\over 3}\int_{1}^\Phi \left(\Phi U_{,\Phi}- 2U\right) d\Phi.
\la{V_U}
\een
Independently of \eqref{V_U}, we suppose that the very point $\bar\Phi=1$ obeys the equation
$V_{,\Phi}(\bar\Phi)=0$.
As a result, we have  $U_{,\Phi}(1)=2$ \cite{Fiziev02}.

The other independent field equations of the system \eqref{MDG_Eq} are defined by its traceless part\footnote{Here
$\hat{X}_\alpha^\beta={X}_\alpha^\beta - {\frac 1 4}X\delta_\alpha^\beta$,
where $X=X_\alpha^\alpha$ is the trace of the tensor ${X}_\alpha^\beta$.}
in the form
\ben
\Phi \hat{R}_\alpha^\beta=-\widehat{\nabla_\alpha\nabla^\beta}\Phi.
\la{MDG_Eq_tr0}
\een

Adding the standard action of the matter fields $\Psi$, based on the minimal interaction with gravity:
\ben
{\cal A}_{matt}={\frac 1 c}\int d^4 x\sqrt{|g|}{\cal L}_{matt}(\Psi,\nabla \Psi; g_{\alpha\beta}),
\la{Amatt}
\een
we obtain for the gravi-dilaton sector the dynamical system
(in cosmological units $\Lambda=1$, $\kappa =1$, $c=1$ which will we use further on everywhere):
\begin{subequations}\label{DGE:ab}
\ben
\Box\Phi+ {\frac 2 3}\big(\Phi U_{,\Phi}(\Phi) -2U(\Phi)\big)= {\frac 1 3} T,\hskip .04truecm \label{DGE:a}\\
\Phi \hat{R}_\alpha^\beta=  -\widehat{\nabla_\alpha\nabla^\beta}\Phi- \hat{T}_\alpha^\beta,\hskip 1.truecm \label{DGE:b}
\een
\end{subequations}
${T}_\alpha^\beta$ being the energy-momentum tensor of the matter.

As seen, the dilaton $\Phi$ does not interact directly with the matter and is a good candidate for the dark matter.
Its interaction with the usual matter goes only trough the gravitational interaction.
In the language of QFT this interaction is a process of second order with respect to
the Newton constant, i.e. a very weak one.

Now we are ready to define the de Sitter vacuum (dSV) as a state in which
\ben
{T}_\alpha^\beta=0,\,\,\,\Phi=\bar\Phi=1:\,\,\,V{,{}_\Phi}(1)=0,\,\,\,\Rightarrow \nonumber\\
{G}_\alpha^\beta=-\delta_\alpha^\beta,\hskip 2.truecm
\la{dSV}
\een
and the Einstein vacuum (EV) as a state in which
\ben
{G}_\alpha^\beta=0,\,\,\,\Phi=\Phi_0:\,\,\,U_{,\Phi}(\Phi_0)=0,\,\,\,\Rightarrow \nonumber\\
{T}_\alpha^\beta+U(\Phi_0) \delta_\alpha^\beta=0.\hskip 1.5truecm
\la{EV}
\een

A remarkable feature of the whole system \eqref{R_Phi}, \eqref{DGE:ab} is its invariance under the local transformations
\ben
\{U,\,T\}\,\,\,\mapsto \,\,\, \{ U+u(x),\,T-4u(x) \}
\la{gauge_transform}
\een
with an arbitrary local field $u(x)$, independent of $\Phi$.

There is one more circumstance that needs a comment.
Actually, Eq. \eqref{DGE:b} is the simplest form of the dynamical equation for the space-time scalar curvature $R$,
not for the dilaton $\Phi$.
Indeed, assuming $U_{,\Phi\Phi}\neq 0$ one obtains from \eqref{R_Phi}
\ben
\Phi= f_{{}_{,R}}(R).
\la{Phi_R}
\een
Here $f_{{}_{,R}}(R)$ is the first derivative of some new function $f(R)$.
Substituting this relation in Eq. \eqref{DGE:b} we obtain it in an explicit form \eqref{gReq:a}.
If in addition $f_{{}_{,RR}}\neq 0$, Eq. \eqref{gReq:a} is equivalent to Eq. \eqref{DGE:a}.
Besides, under the same conditions we can exclude the dilaton $\Phi$ from  Eqs. \eqref{DGE:b}, as well.
Thus, we obtain the new system
\begin{subequations}\label{gReq:ab}
\ben
f_{{}_{,RR}}\Box R + f_{{}_{,RRR}}\left(\nabla R\right)^2- {1\over 3} \left(R f_{{}_{,R}}-2f\right)={1\over 3}\,T, \hskip 0.8truecm \label{gReq:a}\\
f_{{}_{,R}}\hat{R}_\alpha^\beta = - f_{{}_{,RR}}\widehat{\nabla_\alpha\nabla^\beta}R -f_{{}_{,RRR}}\widehat{\nabla_\alpha R\nabla^\beta R}
 -\hat{T}_\alpha^\beta.\hskip 0.8truecm \label{gReq:b}
\een
\end{subequations}

The dynamical equations \eqref{gReq:ab} of the problem look more complicated than the system \eqref{DGE:ab}.
The advantage of the form \eqref{gReq:ab} is in the absence of the dilaton field $\Phi$ in it.
If one solves this system, the field $\Phi$ can be obtained using relation \eqref{Phi_R}.
The disadvantage of the simple form \eqref{DGE:ab} is that it generates a wrong feeling
about possible "independent dynamics" of the dilaton $\Phi$.
This is a little bit subtle circumstance. The character of Eq. \eqref{DGE:a},
which actually describes the dynamics of the scalar curvature $R$,
can be explained in the context of the Legendre transform.
Its specific property is that the basic relations take the simplest possible form,
if one uses a mixed representation in which  half the variables are new and
the other half is the old ones.

The physical equivalence of the above two approaches to the field dynamics
in the gravi-dilaton sector of MDG
needs a careful study. If it takes place, Eqs.
\eqref{DGE:a} and \eqref{gReq:a} present two equivalent forms for description of the additional scalar degree
of freedom which comes into being in both the MDG and $f(R)$ models.
It is frozen in GR where the scalar curvature does not own an independent degree of freedom since $R=T$.
We call this new field degree the scalaron field (see Section \ref{ParticleContent}) 
and the corresponding spinless particle the scalaron \cite{Starobinsky07}.

\section{The Legendre transform which relates MDG and $f(R)$ theories}\label{Legendre}

Relations \eqref{R_Phi} and \eqref{Phi_R} show that the MDG and $f(R)$ theories are related
via the Legendre transform \cite{Arnold89,Zia09}.
Following the traditional notation in gravity,
we obtain some particular form of the Legendre transform.

For equivalence of the MDG and $f(R)$ theories their actions \eqref{A_MDG} and \eqref{Action_fR} must give
equivalent results under the corresponding variations. For this purpose, we need to satisfy the relation
$\mathcal{F}(R,\Phi)= f(R)+2U(\Phi)-R\Phi=\partial_\alpha v^\alpha(x)$ with some vector field $ v^\alpha(x)$
which does not  dependent on $R$ and $\Phi$ {}\footnote{Otherwise the conditions \eqref{restric_dg}
will be destroyed, and in the theory we will have different Noether quantities.}.
It is enough to have  $\mathcal{F}(R,\Phi)= f(R)+2U(\Phi)-R\Phi=\text{const}$.
Then such a vector field $v^\alpha(x)$ certainly exists.

In any case the conditions  $\partial_\Phi \mathcal{F}(R,\Phi)=0$ and $\partial_R \mathcal{F}(R,\Phi)=0$
produce relations \eqref{R_Phi} and \eqref{Phi_R}, respectively,
i.e. the function $\mathcal{F}(R,\Phi)$ generates the Legentre transform.
Adding the condition $\mathcal{F}(R,\Phi)=0$, we obtain the transformation from $U(\Phi)$ to $f(R)$ -- \eqref{Legendre:a},
and the inverse transformation $f(R)$ to $U(\Phi)$ -- \eqref{Legendre:b}, in the following parametric form:
\begin{subequations}\label{Legendre:ab}
\ben
f&=&\! 2\big(\Phi U_{,{}_\Phi}(\Phi) -U(\Phi)\big),\nonumber\\
R&=&2 U_{,{}_\Phi}(\Phi),\,\Phi\in (0,\infty), \hskip .8truecm\label{Legendre:a}\\
U&=&{1\over 2}\big(R f_{,{}_R}(R) -f(R)\big),\nonumber\\
\Phi&=& f_{,{}_R}(R),\,R\in (-\infty,\infty). \hskip .9truecm\label{Legendre:b}
\een
\end{subequations}

The relation $\Phi_{,{R}}R_{,{\Phi}}\equiv 1$ and Eqs. \eqref{Legendre:ab} yield the new ones:
\ben
f_{,{RR}}\,U_{,{\Phi\Phi}}&\equiv& {\tfrac 1 2},\,\,\,f_{,{\Phi}}=2\Phi U_{,{\Phi\Phi}}=\Phi R_{,{\Phi}},\nonumber\\
U_{,{R}}&=&{\tfrac 1 2}R f_{,{RR}}={\tfrac 1 2}R \Phi_{,{R}}, \nonumber\\
f_{,{\Phi}}U_{,{R}}&=&f_{,{R}}U_{,{\Phi}}={\tfrac 1 2}R\Phi,
\la{UPhifR}
\een
which will be used as a dictionary for translation of the results in MDG to f(R)-results and vice versa.

All the above relations are mathematically correct and certainly physically acceptable
if and only if we have convex functions:
$U(\Phi)$ -- in the interval $\Phi\in (0,\infty)$, and $f(R)$ -- in the interval $R\in (-\infty,\infty)$.

Applying the Legendre transform to a function which is not convex,
we will obtain a multi-valued new function. This means that the correspondence
between MDG and f(R) theories will be only local.
To what extend such a local equivalence may be acceptable from a physical point of view
is a problem which we discuss in more details in Section \ref{fR_neq_MDG}.

Hence, there is a one-to-one {\em global} correspondence between the MDG and $f(R)$ models
if and only if
the functions $U(\Phi)$ and $f(R)$ are convex in their physical domains.
Only under this condition the second of the equations Eqs. \eqref{Legendre:a} and \eqref{Legendre:b}
can be solved unambiguously, and the two models are physically equivalent.

\section{The withholding property}\label{WithholdingProperty}

\subsection{The withholding property in the interval $\Phi \in (0,\infty)$}\label{Withholding}

Looking at the dynamical Eq. \eqref{Phi_eq} for dilaton $\Phi(x)=\Phi\left(R(x)\right))$
we see that the only way to force the dilaton to stay all the time in its physical domain
\ben
\Phi \in (0,\infty),
\la{Phi_domain}
\een
preserving each value $\Phi>0$ attainable,
is to impose the conditions
\ben
V(0)=V(\infty)=+\infty
\la{V_cond}
\een
on the dilaton potential $V(\Phi)$.
Then the infinite potential barriers at the ends of the physical domain \eqref{Phi_domain}
will confine dynamically the dilaton $\Phi(x)$ inside this domain,
if it is initially there \cite{Fiziev02,Fiziev03}.
We call this new phenomenon {\em the withholding property} of the dilatonic potential $V(\Phi)$.

For further use we need to derive the consequences of this condition for the cosmological potential $U(\Phi)$.
For a given dilaton potential $V(\Phi)$ under normalization $U(1)=1$ the \eqref{V_U} gives
\ben
U(\Phi)={3\over 2}\Phi^2\!\int_{1}^\Phi \Phi^{-3}V_{,\Phi}d\Phi + \Phi^2.
\la{U_V}
\een

Let us choose a behavior of the dilatonic potential $V(\phi)\sim v\,\Phi^n$ (with some constant $v>0$) to satisfy the condition
\eqref{V_cond}  at the ends of the interval $\Phi\in(0,\infty)$.
Then from Eq. \eqref{U_V} we obtain:
\begin{description}
  \item[a)] For $\Phi \to 0$: $n<0$, and
  \ben
  U(\Phi)\sim {3\over 2}{{|n|}\over{|n|+3}}v\,\Phi^{-|n|-1}.
  \la{Uto0}
  \een
  \item[b)] For $\Phi \to \infty$: $n>0$, and
   \ben
   U(\Phi)\sim
   \begin{cases}
   \,\Phi^2 & \text{for}\,\,\,n\in(0,3),\\
   \,{9\over 2}v\,\Phi^2\ln\Phi  & \text{for}\,\,\,n=3,\\
   \,{3\over 2}{n\over {n-3}}v\,\Phi^{n-1}& \text{for}\,\,\,n>3.
   \end{cases}
   \la{Uto_infty}
   \een
\end{description}

As seen, in each case we have
\ben
U(0)=U(\infty)=+\infty,
\la{U_cond}
\een
but the increase of $U(\Phi)$ and $V(\Phi)$ at the ends of the physical domain \eqref{Phi_domain} is,
in general, not the same.

An observational astrophysical fact:
the cosmological term has a definite positive sign in the observable Universe and
leads to the additional requirement
\ben
U(\Phi)>0\,\,\,\,\text{for}\,\,\,\,\Phi \in (0,\infty).
\la{U_positive}
\een
For potentials with properties \eqref{U_cond}, \eqref{U_positive}
the convex condition reads
\ben
U_{,{}_{\Phi\Phi}}>0\,\,\,\,\text{for}\,\,\,\,\Phi \in (0,\infty).
\la{U_convex}
\een
It ensures the uniqueness of the Einstein vacuum.
We call such $U(\Phi)$ the withholding cosmological potentials.

It is not hard to derive the qualitative behavior of admissible functions $f(R)$ of general type
created via the Legendre transform of the withholding cosmological potential $U(\Phi)$.
Indeed, from Eqs. \eqref{Uto0} and \eqref{Uto_infty}  we obtain the correspondence:
$\Phi\to 0\,\,\,\Leftrightarrow\,\,\,R\to -\infty,\,\,\,f\to -\infty$, and
$\Phi\to +\infty\,\,\,\Leftrightarrow\,\,\,R\to +\infty,\,\,\,f\to +\infty$.
The translation of the properties
\eqref{Phi_domain} and \eqref{U_positive} in the language
of the $f(R)$ models is
$f_{,{R}}>0$ and $f_{,{RR}}>0$ for all $R\in (-\infty,\infty)$.
Hence, f(R) is for sure a strictly monotonically increasing and convex function.

For cosmological potentials $U$ with asymptotic \eqref{Uto0}
and  \eqref{Uto_infty} one obtains from \eqref{Legendre:ab}:
\ben
f(R){\underset{R\to-\infty}{\sim}} -(|n|+2)\left({\frac{3v|n|}{|n|+3}}\right)^{1-\mu^-}\left({\frac{|R|}{|n|+1}}\right)^{\mu^-},\nonumber\\
\mu^-={\frac{|n|+1}{|n|+2}}\in(1/2,1),\hskip 3.9truecm
\la{Rto_pinfty}
\een
\ben
f(R){\underset{R\to+\infty}{\sim}} \begin{cases}
\hskip .truecm {\frac{1}{8}}R^2,\hskip 2.35truecm \text{for}\,\,\, n\in(0,3),\\
\hskip 0truecm 9 v \Phi(R)^2\ln\Phi(R),\hskip .5truecm \text{for}\,\,\, n=3,\\
(n-2)\left({\frac{n-3}{3vn}}\right)^{1/\mu^+ -1} \left({\frac {R}{n-1}}\right)^{1/\mu^+}\\
\hskip 3.3truecm \text{for}\,n>3,
\end{cases}
\la{Rto_minfty}
\een
where in $\Phi(R)=\text{LambertW}\left(\exp\left({\frac{R}{18v}}\right)\right)$ we use the Lambert-W-function \cite{Lambert},
and $\mu^+ = {\frac{n-2}{n-1}}\in(1/2,1)$.

\begin{figure}[htbp]
\centering
\vskip .truecm
\hskip .truecm
\includegraphics[width=8.truecm]{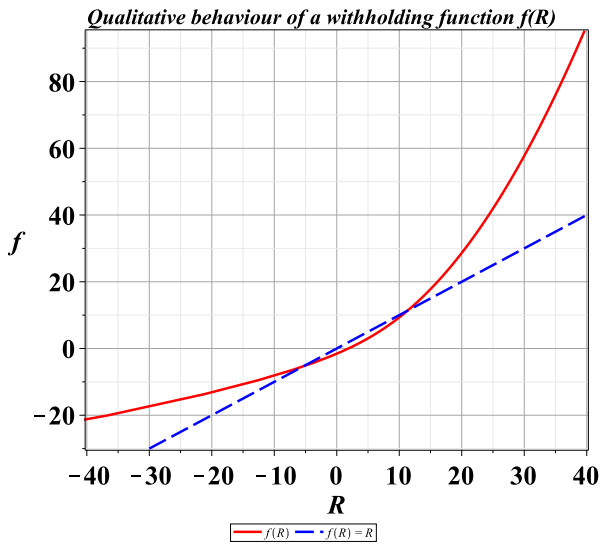}
\vskip -.3truecm
\caption{\small Qualitative behavior of a typical withholding $f(R)$ function}
\label{FigQfR}
\end{figure}

The general form of the function $f(R)$, which follows the asymptotic \eqref{Rto_pinfty} and \eqref{Rto_minfty}
and is in addition a convex function, is shown in Fig.\ref{FigQfR}.
We dub such functions {\em the withholding} $f(R)$ {\em functions}.

\begin{figure}[htbp]
\centering
\vskip .truecm
\hskip .truecm
\includegraphics[width=8.truecm]{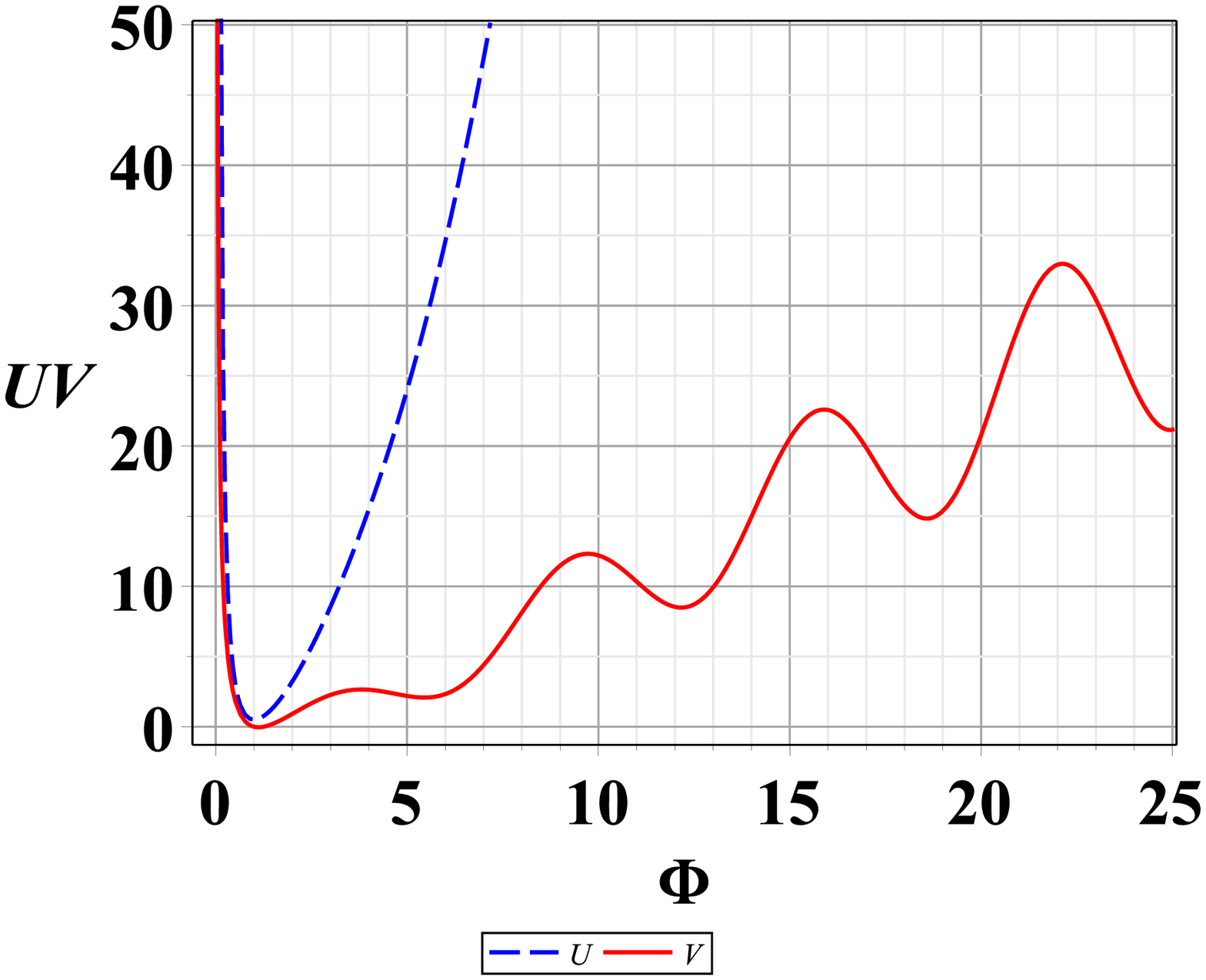}
\vskip -.3truecm
\caption{\small Withholding potentials $U$ (in blue) and $V$ (in red) which have an unique EV and many dSV}
\label{FigV_multivac}
\end{figure}

The uniqueness of the physical de Sitter vacuum in such MDG is still not guaranteed.
Indeed, from Eq. \eqref{V_U} we obtain
\ben
V_{,{}_{\Phi\Phi}}={\frac 2 3}\left(\Phi U_{,{}_{\Phi\Phi}}-U_{,{}_{\Phi}}\right),\,\,\,
V_{,{}_{\Phi\Phi\Phi}}={\frac 2 3}\Phi U_{,{}_{\Phi\Phi\Phi}}.
\la{V_deriv}
\een
These relations show that in the physical domain \eqref{Phi_domain} the functions $U_{,{}_{\Phi\Phi\Phi}}$ and $V_{,{}_{\Phi\Phi\Phi}}$
have the same signs and zeros, but the functions $U_{,{}_{\Phi\Phi}}$ and $V_{,{}_{\Phi\Phi}}$ do not own this property.
Hence, $V(\Phi)$ may have several minima in the domain \eqref{Phi_domain} \cite{Fiziev02}, see Fig. \ref{FigV_multivac}. A similar, but not limited
from below potential with infinite number of minima was considered in a quite different cosmological model in \cite{Steinhardt06}.

Thus, we obtain a new kind of  models with many locally stable dSV at the points $\bar \Phi_{k=0,1,\dots}\geq 1$, $(\bar \Phi_0=1)$
which are solutions of Eq. $\bar \Phi_k U_{\!,{}_\Phi}(\bar \Phi_k)-2U(\bar \Phi_k)=0$ and have
$U_{\!,{}_{\Phi\Phi}}(\bar \Phi_k)> 2U(\bar \Phi_k)/(\bar \Phi_k)^2>0$.
This means that the dilaton behaves around different minima like a field with different masses
$\bar m_{\Phi\,k}^2=(2/3)\left(\bar \Phi_k U_{\!,{}_{\Phi\Phi}}(\bar \Phi_k)- U_{\!,{}_{\Phi}}(\bar \Phi_k)\right)>0$
and excludes a scalar tachyon.
In the language of the $f(R)$ theories we have a series of dSV-curvature-values
$\bar R_k$ defined by Eq. $\bar R_k f{\!,{}_R}(\bar R_k)+f(\bar R_k)=0$ with $f{\!,{}_R}(\bar R_k)/\bar R_k>f{,{}_{RR}}(\bar R_k)>0$.
In each  dSV state  we have different values of the gravitational factor
$\bar G_k=G_N/\bar\Phi_k$ and of the cosmological constant $\bar \Lambda_k=\Lambda U(\bar \Phi_k)\geq\Lambda$.

An interesting questions is whether it is possible to construct
(without introducing more physical structures and fields)
a new mechanism based on dilatonic potentials with many dSV
which is an alternative to the chameleon one and, in addition,
describes a drastic decreasing of cosmological constant
we need for consistency with observations.
\begin{figure}[htbp]
\centering
\vskip .truecm
\hskip .truecm
\includegraphics[width=7.5truecm]{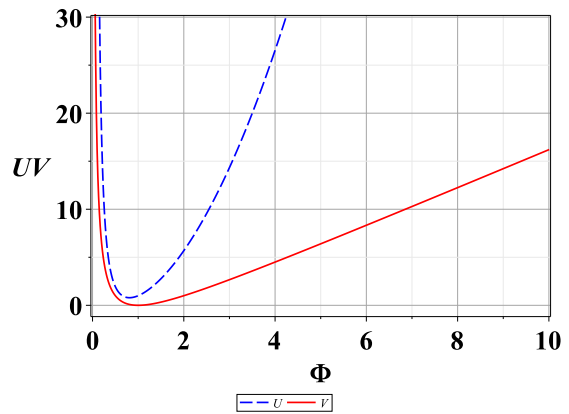}
\vskip -.3truecm
\caption{\small Withholding potentials $U$ (in blue) and $V$ (in red) which have unique EV and dSV}
\label{FigU_V_basic}
\end{figure}

\subsection{Examples of the f(R) models which are equivalent to MDG with unique dSV}\label{fR_equiv_MDG}

There is one more option: To postulate the uniqueness of the dSV \cite{Fiziev02}.
In this case the function $V(\Phi)$ with properties \eqref{V_cond}
will be convex: $V_{\!,{}_{\Phi\Phi}}(\Phi)>0$ for each   $\Phi \in (0,\infty)$. This ensures the stability of dSV
and may exclude a scalar tachyon
since $\bar m^2 =V_{\!,{}_{\Phi\Phi}}(\Phi)=(2/3)\left(U_{\!,{}_{\Phi\Phi}}(1)-2\right)>0$, see Section \ref{ParticleContent}.

Besides, the function
$(2/3)\left(\Phi U_{\!,{}_{\Phi\Phi}}(\Phi)-U_{\!,{}_{\Phi}}(\Phi)\right)=\left(f{\!,{}_R}(R)-Rf{\!,{}_{RR}}(R)\right)/3f{\!,{}_{RR}}(R)>0$
is strictly positive in the whole physical domain and this may avoid the presence of a scalar tachyon in the chameleon models,
see reviews in \cite{Felice10,Sotiriou10,Tsujikawa10,Nojiri11,Rubakov11,Clifton12,Bamba12} and the references therein. In addition the condition $f{\!,{}_{RR}}(R)>0$
entails the requirement
\ben
f{\!,{}_R}(R)/f{\!,{}_{RR}}(R)>R,\,\,\,R\in(-\infty,\infty).
\la{fR_stability}
\een
For $R=\bar R=4$ this yields the condition for stability of dSV \cite{Starobinsky07}.
The stability of dSV in MDG is an obvious consequence of the withholding property.

One can write down a simple example for a pair of withholding potentials with unique dSV:
\ben
V(\Phi)&=&{1\over 2}\mathfrak{p}^{-2}
\left(\Phi+{\frac 1 \Phi} -2\right),\,\,\,\Phi\in(0,\infty),\nonumber\\
U(\Phi)&=& \Phi^2+{3\over {16}}\mathfrak{p}^{-2}
\left(\Phi-{1\over\Phi}\right)^2,
\la{UV_pot}
\een
where $\mathfrak{p}$ is a small parameter, related with the nonzero mass of the dilaton field\footnote{The parameter
$\mathfrak{p}=\sqrt{\Lambda}\hbar/c \bar m$ is the Compton length (measured in cosmological units) of the the dilaton $\Phi$ with mass $m_\Phi=\bar m$.}
\cite{Fiziev02,Fiziev03}, see Fig.\ref{FigU_V_basic}.

The corresponding function $f(R)$ can be
written in the following parametric form:
\ben
f&=&{\frac{3}{8\mathfrak{p}^2}}\left(\Phi^2-{\frac 3{\Phi^2}}+2\right)+2\Phi^2,\nonumber\\
R&=&{\frac{3}{4\mathfrak{p}^2}}\left({\Phi-\frac 1{\Phi^3}}\right)+4\Phi,\,\,\,\,\Phi\in(0,\infty).\hskip .5truecm
\la{fR}
\een

More general potentials with the same property ware described in \cite{Fiziev02} and
can be used as a basis for a very general class of withholding potentials
\footnote{In formula
\eqref{UVgeneral:b} the symbol $\sum^{\bigstar}_{\nu_+,\nu_-}$ means that in the sum we
have to replace  the term $\left(\Phi^{\nu_+ -1}-\Phi^2\right)/ (\nu_+-3)=0/0$ (not well defined for $\nu_+=3$) with $\Phi^2\ln\Phi$.
Similarly, in \eqref{fRn1n2:a} the same symbol denotes the replacement
${\frac{(\nu_+ - 2)\Phi^{\nu_+-1}-\Phi^2}{\nu_+ -3}}= 0/0$ with $\Phi(\ln\Phi+1)$
when $\nu_+=3$,
and in \eqref{fRn1n2:b} -- the analogous replacement
${\frac{(\nu_+ - 1)\Phi^{\nu_+-2}-2\Phi}{\nu_+ -3}}= 0/0$ with $2\Phi(\ln\Phi+1/2)$.
Note also that in the formulae \eqref{UVgeneral:ab}, \eqref{wnunu_cond} and \eqref{fRn1n2:ab}
the discrete summation can be replaced by integration with respect to $\nu_\pm$,
thus obtaining integral representations for a large class of withholding potentials and corresponding functions $f(R)$.}:
\begin{widetext}
\begin{subequations}\label{UVgeneral:ab}
\ben
V(\Phi)&=&{\mathfrak{p}^{-2}}\sum_{\nu_+,\nu_-} {\frac { w_{\nu_+,\nu_-} } {\nu_+ +\nu_-}}
\left({\frac {\Phi^{\nu_+}-1}{\nu_+}}+{\frac {\Phi^{-\nu_-}-1}{\nu_-}}\right),\,\,\,\,\,\nu_\pm>0,\,\,\,\,\Phi\in(0,\infty),\label{UVgeneral:a}\\
U(\Phi)&=& \Phi^2+{{3}\over {2\mathfrak{p}^{2}}}\sum^{\bigstar}_{\nu_+,\nu_-} {\frac { w_{\nu_+,\nu_-} } {\nu_+ +\nu_-}}
\left({\frac {\Phi^{\nu_+ -1}-\Phi^2}{\nu_+-3}}+{\frac {\Phi^{-\nu_- -1}-\Phi^2}{\nu_- +3}}\right).\label{UVgeneral:b}
\een
\end{subequations}
\end{widetext}
By construction $V(1)=0$, $V_{,{}_R}(1)=0$. The additional requirements
\ben
w_{\nu_+,\nu_-}\geq 0,\,\,\,\,\, \sum_{\nu_+,\nu_-} w_{\nu_+,\nu_-}=1
\la{wnunu_cond}
\een
ensure that $U(1)=1$ and $V_{,{}_{RR}}(1)={\mathfrak{p}^{-2}}$,
as well as that the convex condition $U_{,{}_{RR}}(\Phi)>0$ for $\Phi\in(0,\infty)$ is fulfilled.

The corresponding withholding function $f(R)$ is defined in the parametric form:
\begin{widetext}
\begin{subequations}\label{fRn1n2:ab}
\ben
f&=&2\Phi^2 +{\frac{3}{\mathfrak{p}^2}}\sum^{\bigstar}_{\nu_+,\nu_-} {\frac { w_{\nu_+,\nu_-} } {\nu_+ +\nu_-}}
\left({\frac {(\nu_+ - 2)\Phi^{\nu_+ - 1}-\Phi^2}{\nu_+ -3}}-{\frac {(\nu_- + 2)\Phi^{-\nu_- - 1}+\Phi^2}{\nu_- +3}}\right),
\,\,\,\,\nu_\pm>0,\label{fRn1n2:a}\\
R&=&4\Phi +{\frac{3}{\mathfrak{p}^2}}\sum^{\bigstar}_{\nu_+,\nu_-} {\frac { w_{\nu_+,\nu_-} } {\nu_+ +\nu_-}}
\left({\frac {(\nu_+ - 1)\Phi^{\nu_+ - 2}-2\Phi}{\nu_+ -3}}-{\frac {(\nu_- + 1)\Phi^{-\nu_- - 2}+2\Phi}{\nu_- +3}}\right),
\,\,\,\,\Phi\in(0,\infty).\label{fRn1n2:b}
\een
\end{subequations}
\end{widetext}

In all of these cases the necessary and sufficient condition for global equivalence
of the MDG and $f(R)$ models are satisfied.
The functions $U(\Phi)$ and $f(R)$ are both convex and single-valued.
The proper choice of the powers $\nu_\pm$ which enter into the sums can make the withholding barriers
at the ends of the physical domain \eqref{Phi_domain} impenetrable
not only in the classical dynamics of fields under consideration, but also at quantum level.

Much more strong withholding barriers at the ends of the physical domain \eqref{Phi_domain} are produced, for example, by the potential
\ben
U(\Phi)&=&\Phi^2 +{\frac{3}{16\mathfrak{p}^2}} {1\over a} \left(e^{a(\Phi-1/\Phi)^2}-1\right),\nonumber\\
 a&>&0,\,\,\,\Phi\in(0,\infty).
\la{U_exp}
\een
It yields a withholding function $f(R)$ defined  in a parametric form

\ben
f&=&2\Phi^2+{\frac{3}{8\mathfrak{p}^2}}\left(\left(2a\left(\Phi^2-1/\Phi^2\right)-1\right)e^{a(\Phi-1/\Phi)^2}+1\right),\nonumber\\
R&=&4\Phi+{\frac 3{4\mathfrak{p}^2a}}{\frac{\Phi^4-1}{\Phi^3}}e^{a(\Phi-1/\Phi)^2},\,\,\,\Phi\in(0,\infty).
\la{fR_exp}
\een
Its qualitative behavior is the same as the one shown in Fig.\ref{FigQfR}.

From an analytical point of view, when in the formulae \eqref{UVgeneral:ab}, \eqref{fRn1n2:ab}  only
a finite number of terms enters, at the ends of the physical domain \eqref{Phi_domain} we will have:
i) poles;
ii) branching points -- in corresponding functions;
iii) log singularities -- in cosmological potential $U(\Phi)$;
and
iv) more singular term related with the Lambert-W function -- in $f(R)$;
v) the ends of the physical domain are essential singular points,
as in the example \eqref{U_exp}, \eqref{fR_exp}.

\subsection{Direct Construction of withholding functions f(R).}\label{Constuct_fR}

The second relation in Eqs. \eqref{Legendre:a} can be solved explicitly in rare cases.
Even if this is possible, as a rule, the result is complicated and not very useful for further usage.
Taking into account the asymptotics \eqref{Rto_pinfty}, \eqref{Rto_minfty},
and the convex condition we can construct the withholding functions $f(R)$ in a more direct way.

Consider the functions
$$F(x;\mu,\alpha)= \left(x^2+1\right)^{\mu/2} / \left(1+\exp(-2\alpha x)\right)$$
with the parameters $\mu,\alpha$
and construct a five parametric family of $f(R)$ functions
\ben
f(R;\mu_1,\mu_2;\alpha,R_1,R_2)&=&\nonumber\\
=F(R/R_1;1/\mu_1,\alpha)&-&F(-R/R_2;\mu_2,\alpha)\hskip .8truecm
\la{f12}
\een
with the parameters $\mu_{1,2}\in (1/2,1)$, and $\alpha, R_1,R_2 \in (0,+\infty)$.
By construction these functions have the asymptotic behavior \eqref{Rto_pinfty}, \eqref{Rto_minfty}
and satisfy the Starobinsky condition $f(0;\mu_1,\mu_2;\alpha,R_1,R_2)=0$.
The next step is to construct the linear combinations, analogous to Eq. \eqref{UVgeneral:ab}:
\begin{widetext}
\begin{subequations}\label{f_general:ab}
\ben
f(R)=
\sum_{\{\mu_1,\mu_2,\alpha,R_1,R_2\}\in CxD}& W_{\{\mu_1,\mu_2,\alpha,R_1,R_2\}}f(R;\mu_1,\mu_2;\alpha,R_1,R_2),\label{f_general:a}\\
\sum_{\{\mu_1,\mu_2,\alpha,R_1,R_2\}\in CxD}& W_{\{\mu_1,\mu_2,\alpha,R_1,R_2\}}=1,\,\,\,W_{\{\mu_1,\mu_2,\alpha,R_1,R_2\}}\geq 0.\label{f_general:b}
\een
\end{subequations}
\end{widetext}
Unfortunately, the problem of the convex property of these functions is not trivial.

Let us call {\em the convex domain} (CxD) the domain of the parameters $\mu_1,\mu_2,\alpha,R_1,R_2$
for which the functions \eqref{f12} are convex\footnote{Since for the functions \eqref{f_general:a} we have $f(R)\to 0$ for $R\to -\infty$,
and $f(R)\to +\infty$ for $R\to +\infty$, the convex condition $f_{,{}_{RR}}(R)>0$ ensures the property $f_{,{}_R}(R)>0$ for all R.}.
The CxD turns to be not a simple set of points in the five dimensional parameter-space. If we perform the summation
in Eq. \eqref{f_general:a} only over the values $\{\mu_1,\mu_2,\alpha,R_1,R_2\}\in CxD$, then the conditions
\eqref{f_general:b} ensure the convex property of the functions $f(R)$ and these functions are withholding.

Thus, we see one more advantage of the MDG in comparison with the $f(R)$ theories: In MDG the convex domain CxD
has a simple form described in the relations \eqref{UVgeneral:ab} by the inequalities $\nu_\pm>0$.
In contrast, in the $f(R)$ theories  CxD is a quite complicated set of points in the space of the parameters.

\subsection{The withholding property in a finite interval $\Phi \in (\Phi_1,\Phi_2)\subset (0,\infty)$}\label{Nowithholding}
\begin{figure}[htbp]
\centering
\vskip .truecm
\hskip .truecm
\includegraphics[width=7.2truecm]{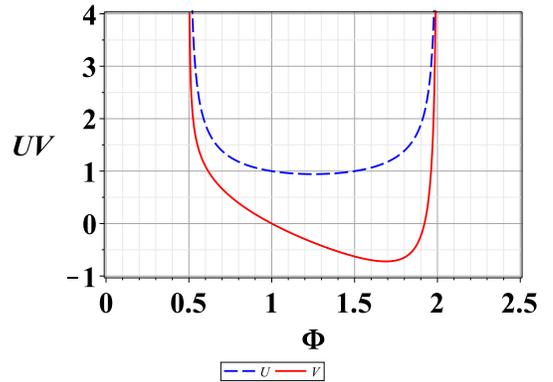}
\vskip -.3truecm
\caption{\small Withholding potentials $U$ (in blue) and $V$ (in red) in the finite domain $(\Phi_1,\Phi_2)=(0.5,2)$ for $\nu_{1,2}=1/2$}
\label{FigPhi1_Phi2}
\end{figure}

Another possibility is to withhold the dilaton field in a finite interval $\Phi\in [\Phi_1,\Phi_2]$ where $0<\Phi_1<1<\Phi_2<\infty$.
We shall not discuss in detail the corresponding class of withholding potentials $U$ and $V$.
The withholding potentials with unique dsV of this class  can be written in the form
$V:=\sum_{\nu_1,\nu_2}w_{\nu_1,\nu_2}V_{\nu_1,\nu_2}$, and
$U:=\sum_{\nu_1,\nu_2}w_{\nu_1,\nu_2}U_{\nu_1,\nu_2}$, $\nu_{1,2}\in (0,1)$,
with coefficients $w_{\nu_1,\nu_2}$ which obey relations \eqref{wnunu_cond}.
The dilatonic potentials $V_{\nu_1,\nu_2}$ have the form
\ben
V_{\nu_1,\nu_2}(\Phi)={\frac {c_{\nu_1,\nu_2}}{\mathfrak{p}^2}}\times \hskip 3truecm \\
\times\Bigg(\!\!\left(\!\left({\frac{1-\Phi_1}{\Phi-\Phi_1}}\right)^{\nu_1}\!\left({\frac{\Phi_2-1}{\Phi_2-\Phi}}\right)^{\nu_2}
\!-1\right)+b_{\nu_1,\nu_2}(\Phi-1)\!\!\Bigg),\nonumber
\la{V_nu1nu2}
\een
where $b_{\nu_1,\nu_2}=\nu_1/(1-\Phi_1)-\nu_2/(\Phi_2-1)$ and $c_{\nu_1,\nu_2}=\left( \nu_1/(1-\Phi_1)^2+\nu_2/(\Phi_2-1)^2\right)^{-1}$,
see an example in Fig.\ref{FigPhi1_Phi2}.
All these potentials satisfy the requirements
$V(1)=0$, $V_{,\,{}_{\Phi}}(1)=0$, and $V_{,\,{}_{\Phi\Phi}}(1)=1/ \mathfrak{p}^2>0$
and contain enough free parameters for fitting real problems.
The potential $U(\Phi)$ can be found using Eq. \eqref{U_V}.

It can be shown that the qualitative behavior of the corresponding function $f(R)$ is the same as shown in Fig.\ref{FigQfR},
but now the asymptotics are
\ben
f(R){\underset{R\to-\infty}{\sim}} \Phi_1 R,\,\,\,\,\,
f(R){\underset{R\to+\infty}{\sim}} \Phi_2 R,\\ 0 < \Phi_1 < \Phi_2 < \infty,\hskip 1.3truecm\nonumber
\la{asymptfR12}
\een
and corresponds to the cases with $\mu_\pm\to 1$ in \eqref{Rto_pinfty}, \eqref{Rto_minfty}.

\subsection{f(R) models which are not equivalent to the MDG.}\label{fR_neq_MDG}

The f(R) models 1 - 10, listed in the Introduction, are not globally equivalent to the MDG.
It is not hard to check that these models suffer from one or several of the following shortcomings:

\begin{itemize}
  \item Negative values of  $\Phi$ are not excluded and the model, being not withholding, contains ghosts.
  \item The model is not withholding because the inadmissible values $\Phi=0$ and/or $\Phi=\infty$ are not excluded.
  \item The model is not withholding since the function $f(R)$ has no necessary asymptotic for $R\to \pm \infty$.
  \item A not convex function $f(R)$ yields multi-valued cosmological potential $U(\Phi)$.

\end{itemize}

These shortcomings do not mean that all these models have no physical significance,
but restrict the physical validity of the corresponding model.
In particular, there may exists cosmologically viable solutions.

For example, the parametric representation of the multi-valued potential $V(\Phi)$
of the astrophysically valuable  Starobinsky 2007 model
can be found in \cite{Frolov08}. It was shown that this potential leads to curvature singularities in
star's models. Nevertheless, for a special initial condition $\Phi_0$ for the dilaton in a tiny domain,
the classical solution $\Phi(t)>0$ is positive for all time instants.
Depending on the choice of the initial conditions the existing singularities in this model
can be not reachable for certain class of classical solutions.
An improvement of the Starobinsky 2007 model was given by Gannouji et al. \cite{Gannouji12}.
This modification has a CxD for proper values of the introduced additional parameter,
thus avoiding multi-valued potentials,
without exclusion of the negative values of the dilaton $\Phi$ for all solutions.

This shows that the proposed in the present paper requirement for a {\em global} equivalence of MDG and $f(R)$ theories is
acceptable but not obligate at classical level\footnote{The author is tankful to A. A. Starobinsly for this remark.}.
However, in the case of a pure local equivalence the $f(R)$ theories are not physically equivalent to MDG.

Another possible scenario to avoid the negative values of $\Phi$ {\em only at the present epoch}
is to choose a proper positive value of $\Phi_0$ and to tune the parameters of the model in a way
which ensures positive $\Phi(t)>0$ for long enough time, i.e. to postpone the disaster for distant future
(in cosmological scales) when the model has to be changed\footnote{The author is thankful to V. A. Rubakov for this remark.}.

In the existing literature we can find also other choices of function $f(R)$.
For example, terms $\sim R^{-n}\left(\ln R/\mu^2\right)^m$,
$\sim\exp(-b R)-1$, or $\sim (\exp(b R)-1)/(\exp(b R)+\exp(b R_0))$ can be included,
see the reviews \cite{Nojiri07,Nojiri11} and the references therein.
In the known to us variety of such functions we were not able to find a $f(R)$ model
which is consistent with the withholding property of MDG.

\section{Presence or absence of ghosts}\label{Ghosts}
In the present paper we call "a ghost" any {\em physical} field with a wrong sign of the kinetic term
in the corresponding lagrangian.
The problem of ghosts in the theories of gravity with higher derivatives has been studied
in many articles, see for example, the articles \cite{Woodard06,Felice06,Faraoni06},
the review articles \cite{Felice10,Sotiriou10,Tsujikawa10,Nojiri11,Rubakov11,Clifton12,Bamba12}
and the references therein. For exclusion of ghosts in the Branse-Dicke theory with nonzero parameter $\omega$
and massless, or very light dilaton, see \cite{Boisseau00}.

\subsection{The conditions for the ghost elimination in MDG}\label{Ghost_MDG}

It is not hard to derive the necessary and sufficient conditions for the absence of ghosts in the MDG.

\begin{enumerate}
  \item There is no ghost-problem for dilaton field since in the action \eqref{A_MDG}
  a kinetic term for $\Phi$ does not exists at all.
  \item  The problem with the presence of ghosts in the metric field $g_{\alpha\beta}$ is not so trivial.
  Generalizing the consideration of Section 93 in \cite{Landau} for the MDG we see that
  in the action \eqref{A_MDG} we will have only kinetic terms of the form
  \ben
  {c\over {8\kappa}}\Phi\, g^{00}\big(\partial_{0} g^{\alpha\beta}\big)^2,\,\,\,\alpha,\beta=1,2,3.
  \la{kinetic_terms}
  \een
  Since $g^{00}>0$, the necessary and sufficient condition for the absence of ghosts in MDG is \eqref{Phi_domain}.
\end{enumerate}

The above results show that the conditions \eqref{Phi_domain} and \eqref{V_cond},
which define the withholding properties,
ensure in a dynamical way also the absence of ghosts in MDG.
Hence, the withholding property is a critical condition for the physical consistency of MDG.

\subsection{Presence of ghosts in f(R) theories of gravity and their relation with MDG}\label{Gosts_fR}

The widespread opinion is that despite the presence of the second derivatives
of the metric $g_{\alpha\beta}$ in the $f(R)$-action \eqref{Action_fR},
there is no problem with ghosts in this modification of GR
(see, for example, the recent paper \cite{Gannouji12}).
As described in some detail in \cite{Woodard06},
the standard reason for this belief is based on the following two steps:

1) Legendre transformation of the $f(R)$-action \eqref{Action_fR} to the MDG-action \eqref{A_MDG}.

2) Conformal transformation of the MDG-action \eqref{A_MDG} to the Einstein-frame-action:
\begin{subequations}\label{Action_EF:abc}
\ben
{\cal A}_{EF}&=&{\frac c {2\kappa}}\int d^4 x\sqrt{|\tilde g|}\tilde R + \hskip .truecm \label{Action_EF:a}\\
&+&2\!\int\! d^4 x\sqrt{|\tilde g|}\left(\!{1\over 2}\tilde g^{\alpha\beta}\partial_\alpha\varphi\partial_\beta\varphi -\tilde V(\phi)\!\right)\!,\nonumber\\
\tilde g_{\alpha\beta}&=&\Phi g_{\alpha\beta},\,\,\,\varphi=\sqrt{3/2}\ln\Phi,\label{Action_EF:b}\\
\tilde V(\phi)&=&e^{-2\sqrt{2/3}\,\varphi}U\!\left(e^{\sqrt{2/3}\,\varphi}\right),\label{Action_EF:c}
\la{Action_EF}
\een
\end{subequations}
which resembles the Hilbert-Einstein-action of GR, plus an additional scalar field $\varphi$
with a normal kinetic term and self-interaction $\tilde V(\varphi)$.
Obviously, ghosts are not present in the Einstein-frame-action \eqref{Action_EF:a}.

However, the action \eqref{Action_EF:a} is produced by the two relations \eqref{Action_EF:b} which both
lose meaning outside the physical domain \eqref{Phi_domain}.
Hence, the accurate analysis shows that the $f(R)$ theories of gravity are not automatically free of ghosts.
For this to be true certain strong additional requirements
equivalent to the above constraints on the MDG are needed.

Some authors mention that for the absence of ghosts in the $f(R)$ theories
one needs the conditions
\ben
f_{,{}_R}(R)>0\,\,\, \text{and}\,\,\, f_{,{}_{RR}}(R)>0.
\la{f_R_f_RR_cond}
\een
However, the natural question:
What kind of physical requirements can ensure the fulfilment of these conditions
and thus indeed can exclude ghosts, remains without a definite answer.
We saw that these conditions do not guarantee the withholding property of the  function $f(R)$, as well as that
the withholding property is critical for exclusion of ghosts.
For this purpose, relations \eqref{f_R_f_RR_cond} are not enough, and
the function $f(R)$ must have  proper asymptotic behavior, say,
described by Eqs. \eqref{Rto_pinfty} and \eqref{Rto_minfty}.

As a result, in Einstein-frame-action
\eqref{Action_EF:abc} the potential $\tilde V(\varphi)$ must have asymptotic
\ben
\tilde V(\varphi)\sim e^{(|n|+3)\sqrt{2/3}\,|\varphi|}\,\,\,\text{ -- for}\quad \varphi\to -\infty,
\la{tildeVaminf}
\een
and for $\varphi\to +\infty$:
\ben
\tilde V(\varphi)\sim
\begin{cases}\,\text{const} & \text{for}\,\,\,n\in(0,3),\\
   \,\varphi  & \text{for}\,\,\,n=3,\\
   \, e^{(n-3)\sqrt{2/3}\,\varphi} & \text{for}\,\,\,n>3.
   \end{cases}
\la{tildeVampinf}
\een

A simple example which corresponds to the cosmological potential \eqref{UV_pot}
is
\ben
\tilde V(\varphi)=1+{3\over {16}}\mathfrak{p}^{-2}\left(1-e^{-2\sqrt{2/3}\,\varphi}\right)^2.
\la{UV_tildeV}
\een

For withholding self-interaction  potentials of general form we obtain
\ben
\tilde V(\varphi)&=& 1+{{3}\over {2\mathfrak{p}^{2}}}\sum^{\bigstar}_{\nu_+,\nu_-} {\frac { w_{\nu_+,\nu_-} } {\nu_+ +\nu_-}}\times \hskip 0truecm\\
&\times&\left({\frac {e^{(\nu_+ -3)\sqrt{2/3}\,\varphi}-1}{\nu_+-3}}+{\frac {e^{-(\nu_- +3)\sqrt{2/3}\,\varphi}-1}{\nu_- +3}}\right).\nonumber
\la{UVgeneral_tildeV}
\een

Another cosmologically valuable example is $\tilde V(\varphi)=V_0\exp(-\sqrt{3}\varphi)$ (in our normalization).
It was considered in \cite{Capozzielo06a}. There it was shown that
this very simple potential offers a good fitting of the observational date
in a large domain, showing some small deviations in the CMBR angular power spectrum for small $l$.
We draw a special attention to this potential,
because it is very close to the withholding property.
One may hope that a small change -- just to obey the withholding property,
may improve the fitting of the known observational data.

\subsection{Particle content of gravi-dilaton sector in MDG and its physics}\label{ParticleContent}

The particle content and the ghost problem of the $f(R)$ theories was discussed as early as in \cite{Muiller88}.

In the present Section we consider the  perturbations around dSV in MDG.
Here the propagators have a more elegant form.
The physical content of the model is more transparent and simpler than in the $f(R)$ theories which are only locally
equivalent to MDG, since a lot of possible cases, considered in \cite{Muiller88}, are automatically excluded in MDG.
In the case of a global equivalence the physical content of the two models is the same.

Consider small perturbation of the metric $g_{\mu\nu}=\bar g_{\mu\nu}+h_{\mu\nu}$
around the solution $\bar g_{\mu\nu}$ for dSV \eqref {dSV}.
The linear approximation for perturbations of Eqs. \eqref{DGE:ab}
above dSV yields the particle content of the gravi-dilaton sector of MDG.
The needed basic geometrical relations are given in the Appendix B. Indeed:

\begin{enumerate}
  \item Let us introduce the scalaron $\zeta$ as excitation of the dilaton above dSV in the following way:
\ben
\zeta = \delta \Phi= \bar f_{\!,{}_{RR}}\delta R={\frac {\bar\Phi}{6}}\,{\frac{\overline{\square} h -2h}{\bar m^2+2\bar U/3\bar\Phi}}.
\la{scalaron}
\een
Then it obeys the scalaron wave equation with a source:
\ben
\big(\hskip .02truecm \overline{\square}+{\tfrac 1 6}\bar R +  m^2\big) \zeta = \delta T/3,\,\,\ \Leftrightarrow\nonumber \\
\zeta ={\frac {1}{\overline{\square}+{\tfrac 1 6}\bar R +  m^2}}\left(\delta T/3\right),
\hskip 0.4truecm
\la{scalaron_Eq}
\een
where the mass of the scalaron, defined as
\ben
m =\sqrt{{\bar m}^2- {\tfrac 1 6}\bar R}\,\,\,\,\,\, \text{for}\,\,\,\,\, {\bar m}^2 \geq {\tfrac 1 6}\bar R
\la{m}
\een
is real, and $\delta T$ is the perturbation of the trace of energy-momentum tensor of matter.
  \item For the traceless part of the metric perturbation $\hat h_{\alpha\beta}$ (which carries the graviton\footnote{Just for simplicity,
  we ignore the detailed spin structure of the graviton. To make it transparent,
  one has to vanish nonphysical gauge degrees of freedom in $\hat h_{\alpha\beta}$.
  This well-known procedure is irrelevant for our proposes and we skip it here.}) we obtain the wave equation:
  \ben
  \left(\overline{\square}+{\tfrac 1 6}\bar R\right)\hat h_{\alpha\beta}={\frac {1}{\bar\Phi}}\left(-\delta \hat T_{\alpha\beta} - \widehat{\nabla_\alpha\nabla_\beta}\zeta\right)
  \,\, \Leftrightarrow  \nonumber \\
  \hat h_{\alpha\beta}={\frac {1}{\overline{\square}+{\tfrac 1 6}\bar R}}\,{\frac {1}{\bar\Phi}}\left(-\delta \hat T_{\alpha\beta} - \widehat{\nabla_\alpha\nabla_\beta}\zeta\right).\hskip .6truecm
  \la{graviton_Eq}
  \een
  $\delta \hat T_{\alpha\beta}$ being the perturbation of the traceless part of energy-momentum tensor of matter.
We see that:
    \begin{itemize}
      \item The propagators of the scalaron (spin $s=0$) and graviton (spin $s=2$)\footnote{The precise meaning of
      these operators in de Sitter space-time is defined by the corresponding two-point functions of different type
      -- symmetric, advanced, retarded, causal, etc. For the first time these were studied in \cite{Chernikov68,Tagirov73}.
      In full detail the explicit form of two-point functions can be found in \cite{Ford85,Mottola85,Allen85}.}:
      \begin{subequations}\label{propagators:ab}
      \ben
      \Pi_{s=0}&=&{\frac {1}{\overline{\square}+{\tfrac 1 6}\bar R+ m^2}},\label{propagators:a}\\
      \Pi_{s=2}&=&{\frac {1}{\overline{\square}+{\tfrac 1 6}\bar R}}\label{propagators:b}
      \een
      \end{subequations}
      look like propagators of a standard classical massive field (scalaron, $s=0$) and massless field (graviton, $s=2$) in de Sitter space-time,
      correspondingly\footnote{The author is tankful to A. A. Starobinsky for his remark about the mass of the graviton:
      It must be zero due to the gauge invariance of the theory with respect to the arbitrary coordinate transformations.}.
      In the graviton case the additional term ${\tfrac 1 6}\bar R$ ensures the conformal invariance of the Eqs. \eqref{graviton_Eq},
      and the zero mass of the graviton.
      \item If ${\bar m}^2 \geq {\tfrac 1 6}\bar R$, the scalaron never becomes a tachyon.
      This inequality can be fulfilled as a result of the withholding property since $\bar m^2 = V{\!,{}_{\Phi\Phi}}(\bar \Phi)>0$
      for $\bar \Phi$ being a point of dilatonic potential minimum.
      We shall see that this condition is met with a large stock,
      since from another reason physically admissible are the values ${\bar m}^2 \gg {\tfrac 1 6}\bar R$.
     \item MDG predicts the existence of free spreading scalaron waves.
     Since the mass of the dilaton is not zero, their velocity is smaller than the velocity of light.
     \item The electromagnetic field and other sources with $T=0$ are not able to create scalaron.
      \item As a result of the withholding property the graviton never becomes ghosts since $\bar \Phi>0$.
      \item The scalaron is a source of gravitational waves.
      This phenomenon is natural for a theory (like GR) in which every physical field curves the space-time
      and thus becomes a source of gravitational waves.
      \item If one tries to express the scalaron contribution in the source of gravitational waves
      directly via $\delta T/3$, one will be forced to use the operator
      \ben
      {\frac {1}{\overline{\square}+{\tfrac 1 6}\bar R}} \widehat{\nabla_\alpha\nabla_\beta}{\frac {1}{\overline{\square}+{\tfrac 1 6}\bar R + m^2}}=\hskip .2truecm\\
      ={\frac {1}{\overline{\square}+{\tfrac 1 6}\bar R}}
      \left(\nabla_\alpha\nabla_\beta-{\frac 1 4}\bar g_{\alpha\beta}\overline{\square}\right)
      {\frac {1}{\overline{\square}+{\tfrac 1 6}\bar R + m^2}}.\hskip -1.95truecm\nonumber
      \la{source_operator}
      \een
      After some manipulations one obtains for the second term in the brackets, enclosed between the two propagators,
      a representation which resembles a presence of a ghost.
      Such interpretation is certainly not correct since this operator is not a propagator but just a source-operator.
      Its physical meaning is clear: the matter creates scalaron waves via $\delta T/3\neq 0$.
      These scalaron waves induce gravitational waves via the source operator \eqref{source_operator}
      when the quadrupole moment is not zero.
    \end{itemize}
\end{enumerate}

Thus, the particle content of the gravi-dilaton sector of MDG and the corresponding basic physics in it is clarified.

\section{Concluding remarks}\label{conclusion}

In the present paper,
we have studied carefully the relation between  Minimal Dilatonic Gravity (MDG) and the f(R) theories of gravity.
These models seem to offer a unified description of dark energy and dark matter.
Our final result is that the two generalizations of GR are globally equivalent under certain additional assumptions.

Their physical equivalence takes place only for certain class of cosmological potentials introduced here and
dubbed withholding potentials since they prevent a change of sign of dilaton $\Phi$,
thus preserving the attractive character of gravity.
It is shown that the same assumptions ensure the absence of ghosts and tachyon in the gravi-dilaton sector of MDG.
Unfortunately, the requirement for withholding property is largely ignored
in the current literature on the $f(R)$ theories.

We studied in detail large classes of withholding cosmological potentials and functions $f(R)$
exploring their asymptotic properties and using the convex condition.
It is shown that the popular choices of the $f(R)$ functions do not own the withholding property and,
as a rule, do not exclude ghosts, being only locally equivalent to MDG.
Some of them suffer also from other shortcomings, discussed in the text.

We develop proper perturbation theory around de Sitter vacuum of MDG.
The background space-time is de Sitter one.
In this framework we defined the scalaron and graviton as a field excitation above the de Sitter vacuum of MDG
and justified their basic physical properties deriving the corresponding wave equations with sources.

According to Eq. \eqref{graviton_Eq}, in MDG graviton is a massless particle in the de Sitter space-time.
In the corresponding conformal-invariant wave operator $\overline{\square}+{\tfrac 1 6}\bar R$
presents a constant term which resembles a mass term in a flat space-time. It corresponds to an extremely small mass, even
in comparison with the electron ones $m_e$:
\ben
m_R\!=\!(\hbar/c)\sqrt{\bar R/6}\!=\!(\hbar/c)\sqrt{2\Lambda/3}\!\approx\! 1.5\times 10^{-38} m_e.
\hskip .4truecm
\la{mg}
\een
This value is consistent with the known astrophysical data about the mass of the graviton \cite{Goldhaber10}.
Such term is needed to have a right quasiclassical limit of the theory, as well as to have a unique vacuum
(up to a unitary equivalence) in the Fock
space in QFT in the de Sitter space-time \cite{Chernikov68,Tagirov73,Ford85,Mottola85,Allen85,Tagirov05}.

The mass of the scalaron $m$
in Eqs. \eqref{scalaron_Eq}, \eqref{m} may be arbitrary large.
The only known restriction, needed to avoid conflicts with the solar system and laboratory
experiments was pointed out in the Introduction.
It gives
\ben
m \gtrapprox 2\times 10^{-9} m_e\approx  10^{29} m_{R}.
\la{mPhi}
\een

Thus, the second condition in Eq. \eqref{m} is certainly fulfilled in realistic MDG models.
As seen, in their gravi-dilaton sector we have two very different mass scales.

We recover two new phenomena in MDG:
\begin{enumerate}
  \item Scalaron waves freely spreading in vacuum with velocity
much smaller than the velocity of light
because of relation \eqref{mPhi}.
This is consistent with the expected low velocity of dark matter particles
and supports the hypothesis that the scalaron may be a good candidate for dark matter.
An important problem is to find experimental methods and/or observational tools
for a direct registration of scalaron waves predicted by MDG.

Our consideration shows that from a geometrical point of view
the scalaron waves are freely spreading in vacuum perturbations
of space-time scalar curvature $R$ (See Eq. \eqref{scalaron}.).
Thus, their existence is an essential deviation from GR.
Here the scalar curvature is rigidly related with the
trace of the energy-momentum tensor of matter and vanishes
identically in vacuum.
  \item Induction of gravitational waves by the scalaron, according to Eq. \eqref{graviton_Eq}.
Estimates of the magnitude of this effect related with astrophysical objects
and with the early Universe are needed to see whether this prediction of MDG
may be of interest for the future more sensitive detectors of gravitational waves.
\end{enumerate}

According to the withholding condition,
the dilatonic potential has at least one minimum.
Hence, in MDG there exists at least one dSV.
If there are several of them, as shown in Fig. \ref{FigV_multivac},
then the withholding condition ensures the ordering $0<\bar \Phi_0<\bar \Phi_1< \dots <\infty$.

Our physical normalization requires for the lowest minimum $\bar \Phi_0=1$ and $U(\Phi_0)=1$ which is
a natural choice for the case of unique dSV in MDG \cite{Fiziev02,Fiziev03}.
If there exist many dSV, this normalization corresponds to the hypothesis that the present state of
the Universe is close to the lowest one.

For different dSV, numbered by $k=0,1,\dots$, the gravitational factor has different values $\bar G_k=G_{N}/\bar \Phi_k$.
The different values of the cosmological factor are $\bar \Lambda_k = \Lambda \bar U_k$.
Hence, as a result of the withholding property,
the lowest dSV is a state with minimal value of the cosmological factor
and with maximal value of the gravitational factor.
The scalaron has also different masses $m_k$ in the vicinity of different dSV.
This new physical situation needs a more careful analysis.

A basic open problem remains a careful choice of a withholding MDG model
which is able to fit known observational data in cosmology and astrophysics.
The present paper is a necessary step in this direction which outlines
a variety of possible future developments.

\begin{acknowledgments}
The author is deeply indebted to the Directorate of the Laboratory of Theoretical Physics, JINR, Dubna for the good
working conditions during his stay there in 2010-2012.

He is grateful to D.V. Shirkov, B. M. Barbashov, D.I. Kazakov, A.S. Sorin and V.V. Nesterenko for useful
discussions and for raising stimulating questions.

The discussions with V.A. Rubakov, T.P. Sotiriou, and their comments were extremely helpful for the
presented here new development of the topic.

The author is thankful to S. D. Odintsov for drowing attention to the articles \cite{Nojiri03,Nojiri04,Abdalla05,Nojiri08c,Cognola08,Nojiri07,Nojiri11,Bamba08}.

The author wishes to express his sincere gratitude to A.A. Strarobinsky for critical reading of the first version of this article
and many important suggestions and references.

The author is thankful also to unknown referee for his useful comments and suggestions about reduction of the initial text of the manuscript.

This research was supported in part by the Foundation for Theoretical and Computational Physics and Astrophysics, and
by a Grant of the Bulgarian Nuclear Regulatory Agency for 2012.

\end{acknowledgments}

\appendix

\section{Derivation of the MDG-field-equations}

In this article, we use the notation written in the sign conventions of the article  \cite{Starobinsky07}. Then
\ben
\delta_g \left(\sqrt{|g|}R\right)&=&\sqrt{|g|}\bigg(G_{\alpha\beta}\delta g^{\alpha\beta}+\nabla_\lambda \left(\delta v^\lambda\right)\bigg),\nonumber\\
G_{\alpha\beta}&=&R_{\alpha\beta}-{1\over 2}g_{\alpha\beta}R.\hskip .truecm
\la{varg2}
\een
The explicit form of the vector  $\delta v^\lambda$ is not needed in GR.
The direct calculations give
\ben
\delta v^\lambda= g^{\lambda\nu}\delta_g\Gamma_{\mu\nu}{}^\mu -g^{\alpha\beta}\delta_g\Gamma_{\alpha\beta}{}^\lambda=\nonumber \\
=\partial_\nu g^{\lambda\nu}+\Gamma_{\alpha\beta}{}^\lambda\delta g^{\alpha\beta} +
\Gamma_\nu\delta g^{\lambda\nu}+2 g^{\lambda\nu}\delta_g \Gamma_\nu
\la{dv}
\een
and after some more algebra we obtain
\ben
\sqrt{|g|} \nabla_\lambda\Phi\delta v^\lambda=\sqrt{|g|}\bigg(\nabla_\alpha\nabla_\beta\Phi-g_{\alpha\beta}\Box \Phi\bigg)\delta g^{\alpha\beta}-\nonumber\\
-\partial_\nu\bigg( \sqrt{|g|}\nabla_\lambda\Phi\left(\delta g^{\lambda\nu} -2 g^{\lambda\nu}\delta\left(\ln\sqrt{|g|}\right)\right)\bigg).\hskip .5truecm
\la{x}
\een

For derivation of the equations  \eqref{MDG_Eq}, the variations of the metric coefficients $\delta g^{\lambda\nu}$
and the variations of their derivatives $\delta \left(\partial_\nu g^{\lambda\nu}\right)$
must obey the system of four restrictions on the surface $\partial V^{(3)}$:
\ben
\Phi \delta v^\lambda -\nabla_\nu\Phi\,\delta g^{\lambda\nu} +g^{\lambda\nu}\nabla_\nu\Phi\,\delta\left(\ln|g|\right)=\nonumber \hskip 1.5truecm\\
= \delta\left(\partial_\nu g^{\lambda\nu}\right)+\partial_\nu\left(\ln(\Phi\sqrt{|g|})\right)\delta g^{\lambda\nu}
+\Gamma_{\alpha\beta}{}^\lambda\delta g^{\alpha\beta}+\nonumber\\
+g^{\lambda\nu}\Big( \partial_\nu\left(\delta \ln|g|\right)- \left(\partial_\nu\Phi\right)\delta \ln|g|\Big)=0,\hskip 1.8truecm
\la{restric_dg}
\een

\section{Geometrical perturbations of the de Sitter space-time}
We use the general formalism for perturbations of the metric  $g_{\mu\nu}=\bar g_{\mu\nu}+h_{\mu\nu}$ on a curved
background space-time with metric $\bar g_{\mu\nu}$ \cite{Landau}, corrected with respect to our conventions.
The perturbation $h_{\mu\nu}=g_{\mu\nu}-\bar g_{\mu\nu}=\delta g_{\mu\nu}$ is a small quantity of the first order.
The sign $\approx$ denotes equalities valid in the linear approximation.
The bar sign denotes the phone-space-time-quantities.

The tensor of the affine deformation is
\ben
\delta\Gamma_{\alpha\beta}{}^\gamma\approx H_{\alpha\beta}{}^\gamma={\tfrac 1 2} \left(\overline{\nabla}_\alpha h_\beta^\gamma+\overline{\nabla}_\beta h_\alpha^\gamma-
\overline{\nabla}^\gamma h_{\alpha\beta}\right).
\la{deltaGamma}
\een
Then
\ben
\delta R_{\alpha\beta\gamma}{}^\delta\approx -2\,\overline{\nabla}_{[\alpha} H_{\beta]\gamma}{}^\delta,
\la{deltaRieman}
\een
and imposing the transversal gauge condition
\ben
\overline{\nabla}_\mu h_\alpha^\mu-{\tfrac 1 2} \overline{\nabla}_\alpha h=0,\qquad h=h_\mu^\mu,
\la{transversal}\label{deltaRicci:ab}
\een
from Eqs. \eqref{deltaRieman} and \eqref{transversal} one obtains the general form of the variations of the Ricci tensor and scalar
\begin{subequations}\label{deltaRicciT:ab}
\ben
\delta R_{\alpha\beta} &\approx& {\tfrac 1 2}\, \overline{\square} h_{\alpha\beta}+ h^\sigma_{ \{ \alpha} \bar R^{}_{\beta \}\sigma}-
 h^{\mu\sigma}\bar R_{\mu\alpha\beta\sigma}, \hskip .5truecm\label{deltaRicci:a}\\
\delta R &\approx& {\tfrac 1 2} \overline{\square} h - h^{\mu\sigma}\bar R_{\mu\sigma}.\label{deltaRicci:b}
\een
\end{subequations}

Taking into account the definition of the dSV in MDG it is easy to obtain the following dSV-values:
\begin{subequations}\label{dSVR:abc}
\ben
\bar\Phi :\,\, V_{\!,{}_\Phi}(\bar\Phi)&=&0,\,\, V_{\!,{}_{\Phi\Phi}}(\bar\Phi)=\bar m^2>0,\label{dSVR:a}\\
\bar U &=& U(\bar\Phi),\,\,\bar U_{\!,{}_\Phi}=2\bar U/\bar\Phi,\nonumber \\
\bar U_{\!,{}_{\Phi\Phi}}&=&2\bar U/\bar\Phi^2+{\tfrac 3 2}\bar m^2/\bar\Phi,\label{dSVR:b}\\
\bar R&=&4,\,\,\bar R_{\alpha\beta}=\bar g_{\alpha\beta},\nonumber\\
\bar R_{\alpha\beta\gamma\delta}&=&-{\tfrac 1 3}\left(\bar g_{\alpha\gamma} \bar g_{\beta\delta}- \bar g_{\alpha\delta} \bar g_{\beta\gamma}\right).\label{dSVR:c}
\een
\end{subequations}

The corresponding values for $\bar f$, $\bar f_{\!,{}_{R}}$, and $\bar f_{\!,{}_{RR}}$ can be obtained using Eqs. \eqref{Legendre:ab}.

As a result of Eqs. \eqref{dSVR:abc} we obtain the following final form of the perturbations above dSV-space-time:
\begin{subequations}\label{deltaRicci_dSV:abcde}
\ben
\delta R_{\alpha\beta} &\approx& {\tfrac 1 2}\, \overline{\square}\,\hat{h}_{\alpha\beta}+{\tfrac 4 3}\,\hat{h}_{\alpha\beta}
+{\tfrac 1 8}\, \overline{\square} h\, {\bar g}_{\alpha\beta}, \hskip .5truecm\label{deltaRiccii_dSV:a}\\
\widehat{\delta R_{\alpha\beta}} &\approx& {\tfrac 1 2}\, \overline{\square}\,\hat{h}_{\alpha\beta}+{\tfrac 4 3}\,\hat{h}_{\alpha\beta}=
\delta\!\left(\hat R_{\alpha\beta}\right) +\hat h_{\alpha\beta},\hskip .5truecm \label{deltaRiccii_dSV:c}\\
\delta\!\left(\hat R_{\alpha\beta}\right)&\approx&{\tfrac 1 2}\,\overline{\square}\,\hat{h}_{\alpha\beta}+{\tfrac 1 3}\,\hat{h}_{\alpha\beta}=
{\tfrac 1 2}\,\left(\overline{\square}+{\tfrac 1 6}\bar R\right)\hat{h}_{\alpha\beta},\hskip .8truecm \label{deltaRiccii_dSV:d}\\
\delta R &\approx& {\tfrac 1 2}\, \overline{\square} h -h.\label{deltaRiccii_dSV:e}
\een
\end{subequations}

The used barred quantities represent the corresponding values for any possible dSV.
To simplify the notation, here we do not number explicitly the different dSV.


\end{document}